\documentclass[twocolumn]{aastex6}

\usepackage{amsmath}
\slugcomment{accepted for publication, AJ}

\shorttitle{A Polar with Absorption-Emission Line Reversals}
\shortauthors{Littlefield et al.}

\begin{document}

\title{MASTER OT J132104.04+560957.8: A Polar with Absorption-Emission Line Reversals}
\author{Colin Littlefield,\altaffilmark{1} Peter Garnavich,\altaffilmark{1} Taylor J. Hoyt,\altaffilmark{2} Mark Kennedy\altaffilmark{1, 3}}
\altaffiltext{1}{Department of Physics, University of Notre Dame, Notre Dame, IN 46556}
\altaffiltext{3}{Department of Astronomy \& Astrophysics, University of Chicago, Chicago, IL 60637}
\altaffiltext{3}{Department of Physics, University College Cork, Cork, Ireland}

\begin{abstract}

We present time-resolved photometry and spectroscopy of the recently classified polar MASTER OT J132104.04+560957.8. The spectrum shows a smooth, non-thermal continuum at the time of maximum light, without any individually discernible cyclotron harmonics. Using homogeneous cyclotron modeling, we interpret this as cyclotron radiation whose individual harmonics have blended together, and on this basis, we loosely constrain the magnetic field strength to be less than $\sim$30 MG. In addition, for about one-tenth of the orbital period, the Balmer and He I emission lines transition into absorption features, with He II developing an absorption core. We use our observations of this phenomenon to test theoretical models of the accretion curtain and conclude that the H and He I lines are produced throughout the curtain, in contravention of theoretical predictions of separate H and He I line-forming regions. Moreover, a significant amount of He II emission originates within the accretion curtain, implying that the curtain is significantly hotter than expected from theory. Finally, we comment on the object's long-term photometry, including evidence that it recently transitioned into a prolonged, exceptionally stable high state following a potentially decades-long low state.

\end{abstract}



\section{Introduction} \label{intro}

\subsection{Magnetic Cataclysmic Variables}

Cataclysmic variable stars (CVs) are interacting binaries in which a white dwarf (WD) accretes from a companion star that overfills its Roche lobe. Polars are a type of CV featuring a WD whose magnetic field is sufficiently high  \citep[$\gtrsim 10$ MG;][] {cropper} to synchronize its own rotational period with the binary orbital period. The accretion flow from the donor star will follow a ballistic trajectory until its ram pressure is matched by magnetic pressure from the WD. Moving out of the system's orbital plane, the flow then travels along the WD's magnetic field lines until it slams into a cyclotron-emitting accretion region near one of the WD's magnetic poles. Thus, no accretion disk can form, unlike non- or weakly magnetic CVs. The magnetically channeled part of the accretion flow is commonly referred to as an accretion funnel or an accretion curtain, and the region in which the accretion flow is diverted from its ballistic trajectory is known as the threading region. \citet{cropper} and \citet{hellier} both provide detailed descriptions of polars.

Polars have several important observational characteristics resulting from the high magnetic moment of the WD. For example, the spectrum of a typical polar is dominated by single-peaked emission lines, including high-ionization lines. Intense, single-peaked He II $\lambda\ 4686$ \AA\ is a common indicator of magnetic accretion. Moreover, cyclotron emission from the accretion region can contribute a significant fraction of the system's overall optical light. Because cyclotron radiation is highly beamed, its contribution to the light curve can be highly variable as our viewing angle of the accretion region changes across the orbital period. Thus, the orbital light curves of polars can exhibit rapid, high-amplitude variations seldom seen in non-magnetic CVs.

\subsection{MASTER OT J132104.04+560957.8}

\begin{figure*}[ht!]
\epsscale{1.17}
\plotone{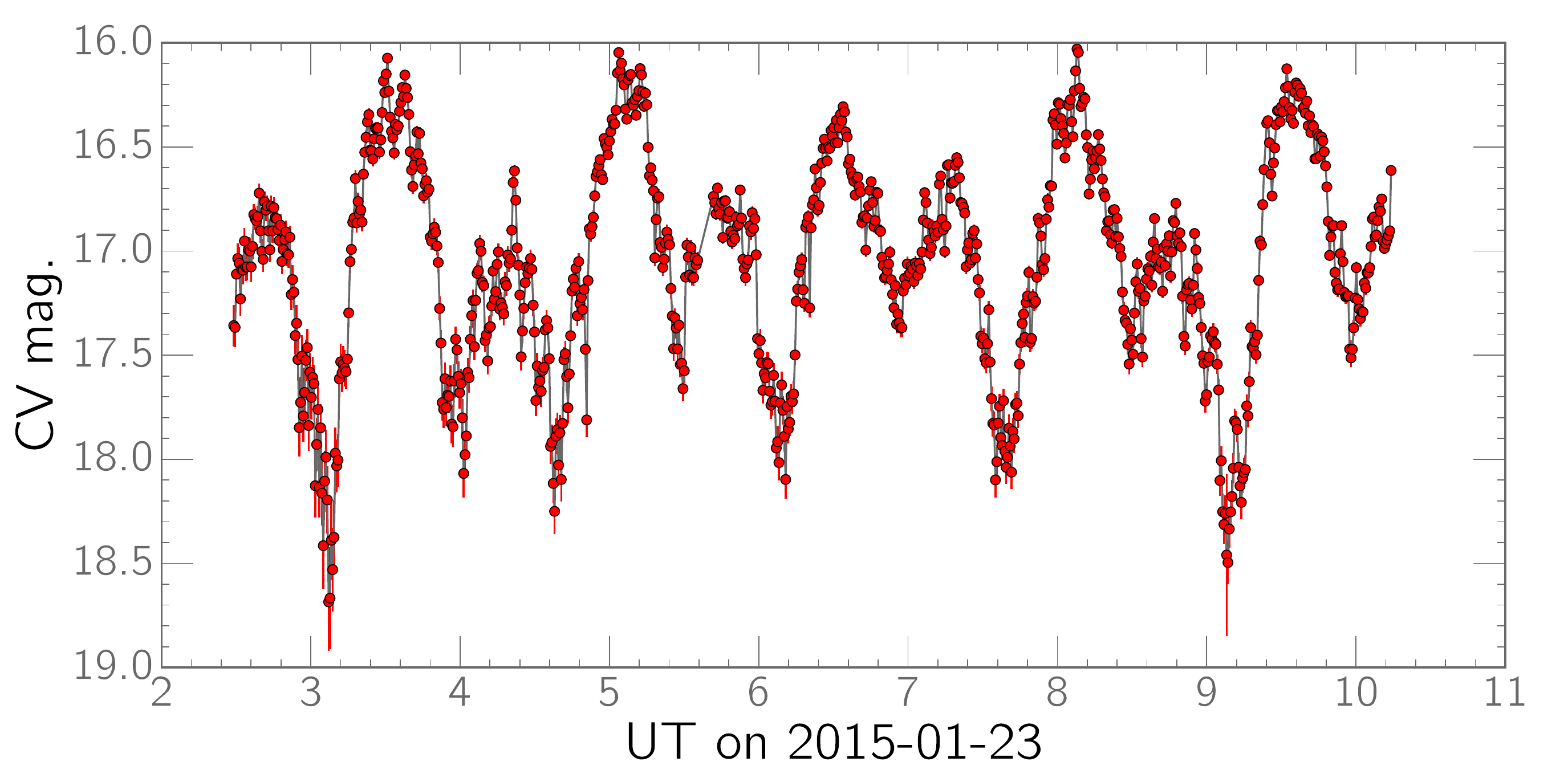}
\caption{A long light curve of J1321 obtained with Van Vleck Observatory's 60-cm Perkin Telescope, illustrating the overall stability of the double-humped structure. Nevertheless, there are noticeable cycle-to-cycle variations, including the changes in the depth of the minima. The data are unfiltered with an exposure time of 30 sec, and the overall cadence is $\sim32$ sec per image. \label{LC}}
\end{figure*}

MASTER OT J132104.04+560957.8 (hereinafter, J1321) was first detected as an optical transient by \citet{Y12} and was subsequently found to exhibit 2-magnitude optical variations with an approximate period of 1.5 hours \citep{kato}. \citet{L15} published the first spectral and photometric observations of J1321 and classified the system as a likely polar based on the detection of strong, single-peaked H, He I, and He II $\lambda\ 4686$ \AA\ emission lines. Their average spectrum revealed a curiously red continuum that peaked near 540~nm without showing any spectral features from the donor star. 

The phase plot of their photometry consisted of two prominent, partially overlapping humps, each of which persists for $\sim~35$\% of the orbit. However, in several instances, the fainter of the two humps displayed extraordinary morphological variations across consecutive orbital cycles (see Fig.~ 2 in \citealt{L15}). The exact cause of the variation was never pinpointed, but it was suggestive of significant changes in the accretion geometry on timescales of just 90 minutes.

\section{New Observations}

\subsection{Simultaneous LBT Spectroscopy \& Photometry}

We obtained simultaneous spectroscopy and photometry of J1321 using the Large Binocular Telescope (LBT) in binocular mode with the Multi-Object Dual Spectrograph \citep[MODS;][]{pogge10} and the Large Binocular Camera \citep[LBC;][]{giallongo08}. The LBT observations started 2015 May 19.367 (UT) and lasted 2.1 hours, or about 1.4 orbits of J1321.

Thirty-five spectra were obtained with MODS1 (using the SX mirror) in the grating mode using a 1.0-arcsec slit. Each exposure was 120~s, and read-plus-overhead time was 100~s. The blue channel covers the wavelength range from 320~nm to 562~nm, and the red channel records 567~nm to 1000~nm. The gap in wavelength coverage is caused by a dichroic filter that divides the light into the two spectrographs.

The spectra were bias-subtracted and flat-fielded using standard reduction procedures. The two-dimensional spectra were extracted using the {\sc 2dspec} package in IRAF. Wavelength calibration was performed using an argon lamp on the blue side and a neon lamp for the red. The spectra were transformed to flux using spectra of the spectrophotometric standard star HZ~44 obtained on the same night.

As MODS1 was taking spectra using the SX mirror, the LBC-red camera was recording images with the DX mirror. LBC is a mosaic of four CCDs, but only 1000 rows of chip~2 were read to reduce overheads. An exposure time of 15.2~s combined with a readout of 20~s resulted in 204 images taken simultaneously with the MODS1 spectra. The seeing over the time series varied between 1.1~arcsec and 1.3~arcsec (FWHM). All of the images were taken through a Bessel $V$ filter, and relative photometric uncertainties for J1321 are 5~millimag.

\subsection{Additional Photometry}
\label{photometry_sec}

Between 2013 and 2017, we obtained additional time-resolved photometry with the 60-cm Perkin Telescope at Wesleyan University's Van Vleck Observatory, the 80-cm Sarah L. Krizmanich Telescope (SLKT) at the University of Notre Dame, and the 1.8-m Vatican Advanced Technology Telescope (VATT).\footnote{Some of the VATT data were previously reported in \citet{L15}.} The exposure time for the first two instruments was 30~s, with between 2 and 4~s of overhead between consecutive images. For the VATT, exposure times were generally 20~s with 12~s of readout, but for one run, we used the Galway Ultra Fast Imager to achieve 10~s exposures with negligible overhead. The Perkin and SLKT time series were unfiltered in an effort to achieve a reasonable signal-to-noise ratio; we inferred approximate $V$ magnitudes by using the $V$ magnitude of the comparison star when calculating J1321's $V$ magnitude, a bandpass which we denote in our figures as ``CV.'' The VATT data used a $V$ filter. Figure~\ref{LC} shows a representative light curve from the Perkin Telescope, covering five consecutive orbital cycles.

\begin{figure}
\epsscale{1.17}
\plotone{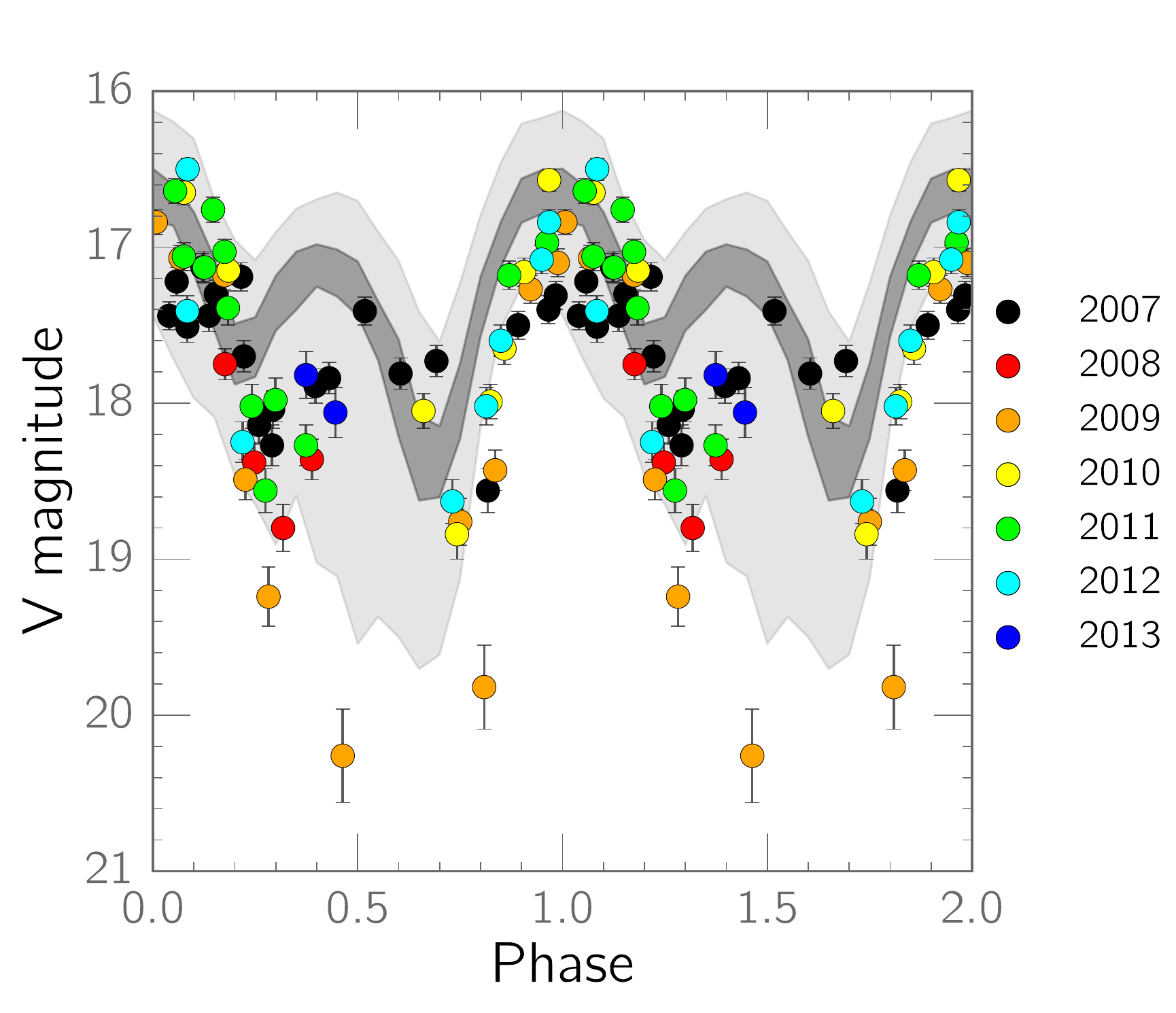}
\caption{A seven-year phase plot of the CRTS photometry using the ephemeris in Sec.~\ref{photometry}, illustrating the long-term stability of the light curve between 2007 and 2013. To enable a direct comparison with our time-series photometry (Sec.~\ref{photometry_sec}), we plot contours of the phase plot of our data. The dark and light contours indicate the innermost 50\% and 98\% of our data at a given phase, respectively. The amplitude of the primary maximum appears to have been lower in 2007 compared to our recent time series. The three faintest observations occur at phases consistent with missing secondary maxima (see text). \label{crts}}
\end{figure}

\subsection{Availability of Data}

Our full photometric and spectroscopic dataset is freely available as supplemental online material, but we request that this work be cited should these data be used in a subsequent work.

\section{Analysis}

\subsection{Long-Term Photometric Behavior}

Our observations and those in \citet{L15} show J1321 to be in a stable high state, consistent with a stable mass-transfer rate. However, archival photometry shows that J1321 underwent a period of prolonged dormancy before transitioning into a high state at some point between 2003 and 2007. Images taken in 1953 (POSS-I O and POSS-E Red), 1983 (Quick-V Northern), 1991 (POSS-II N), 1993 (POSS-II Red), and 1994 (POSS-II Blue) show the system in a very faint state, as do SDSS images from 2003 March, when $g = 21.65$. However, between 2007 and 2013, the Catalina Real-Time Transient Survey \citep[CRTS;][]{drake} observed J1321 a total of 65 times, and during that span, it was detected in a high state exclusively. To illustrate the stability of this high state, Figure~\ref{crts} plots the CRTS data, phased according to our ephemeris. Although the CRTS data in 2007 reveal the primary photometric maximum to have been several tenths of a magnitude fainter in 2007, the observed scatter at any given orbital phase in the remainder of the data is consistent with observed cycle-to-cycle variations. Thus, the CRTS data imply that there were no major variations in the overall brightness between 2007 and 2013, and our own monitoring of J1321 show that its overall brightness has remained unchanged since 2013. Because variations in the overall brightness are directly indicative of variations in the mass-transfer rate, it follows that J1321's mass-transfer rate has been very stable since at least 2007. 

\citet{ms17} studied 44 well-sampled light curves of polars observed by CRTS over a span of $\sim$10 years, and their findings provide a useful context for assessing J1321's long-term light curve. The polars in their sample show an eclectic range of behaviors, with some remaining in persistent low states, others remaining locked in bright states, and still others alternating between the two. Were J1321 included in their sample, it would clearly be classified as a high-state polar, but none of the polars examined in their paper shows a transition similar to J1321's dramatic emergence from an extended faint state into a prolonged, apparently uninterrupted high state. However, this behavior in J1321 becomes apparent only when CRTS photometry is examined alongside older data, so it is possible that some polars in their study might have undergone similar changes before CRTS coverage commenced.

\subsection{Time-Series Photometry}\label{photometry}

\begin{figure*}
\epsscale{1.17}
\plotone{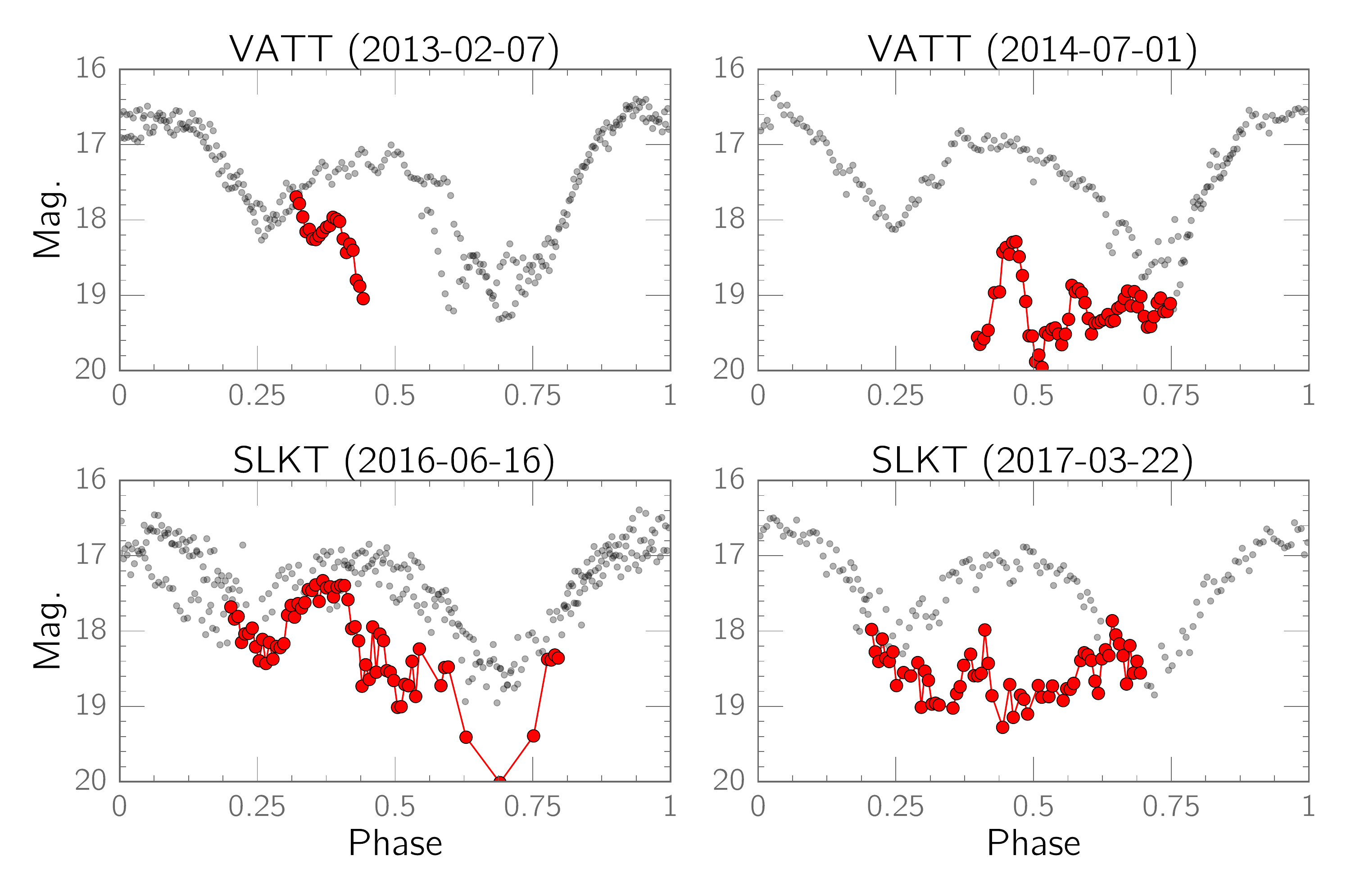}
\caption{Four nightly light curves showing ``missing'' secondary maxima, presented as phase plots to underscore how the missing maxima do not affect the rest of the light curve. The measurements showing the missing maxima are plotted in red. In the top-left panel, the missing maximum began at the end of the observations, and in the top-right panel, it was already underway at the start of the time series. In the bottom-left panel, the three faintest measurements are 4.5-minute bins, as J1321 was undetectably faint in individual images. That particular missing maximum shows that the system began to rise toward secondary maximum as usual between $\phi \sim 0.3-0.4$, but it then faded precipitously in comparison to the two previous cycles. 
\label{secmax}}
\end{figure*}

J1321's photometric behavior during our observations was nearly identical to the general behavior described in \citet{L15}. Each photometric cycle remained dominated by two partially overlapping humps of unequal amplitude, along with two minima of different depths. The brighter of the two humps is typically one-half of a magnitude brighter than the secondary maximum.

To generate an updated ephemeris, we fit second-order polynomials to each of the observed photometric maxima and performed an O$-$C analysis on the measured times of maxima. We find that the photometric maximum follows an ephemeris of $$ T_{max}(BJD) = 2456284.8043(11) + 0.063235199(40) \times\ E.$$ The numbers in parentheses are the $1\sigma$ uncertainties on the final two digits, as determined from Monte Carlo simulations. Given J1321's classification as a polar, we presume that this period corresponds with both the binary orbital period and the WD's rotational period.

\subsubsection{``Missing'' Secondary Maxima}

Although J1321's overall behavior was consistent from night-to-night, we did observe two ``missing'' secondary photometric maxima, not including those previously reported in \citet{L15}. As noted in \citet{L15}, the secondary maximum is sporadically absent from a photometric cycle, essentially disappearing and reappearing on timescales of one orbit. While the secondary maximum usually peaks around V$\sim17$, the system drops below V$\sim$19 during a ``missing'' secondary maximum, and the customary hump-like appearance of the secondary maximum is replaced by irregular variation. Of the 45 observed secondary maxima in our data, only four are missing.\footnote{Another light curve, obtained on 30 June 2014 and shown in Fig. 2 of \citet{L15}, appears to show the very end of a fifth missing secondary maximum, but the data do not cover enough of the event to establish it conclusively as a missing maximum.} Figure ~\ref{secmax} plots these four light curves.

The missing secondary maxima do not appear to be correlated with any concomitant changes in the orbital light curve. In particular, there are no anomalies in the appearance of the primary maximum before or after a missing secondary maximum. This could indicate that the two photometric maxima are associated with different magnetic poles and that the accretion flow to one pole is less stable than the other.

\begin{figure*}
\epsscale{1.17}
\plotone{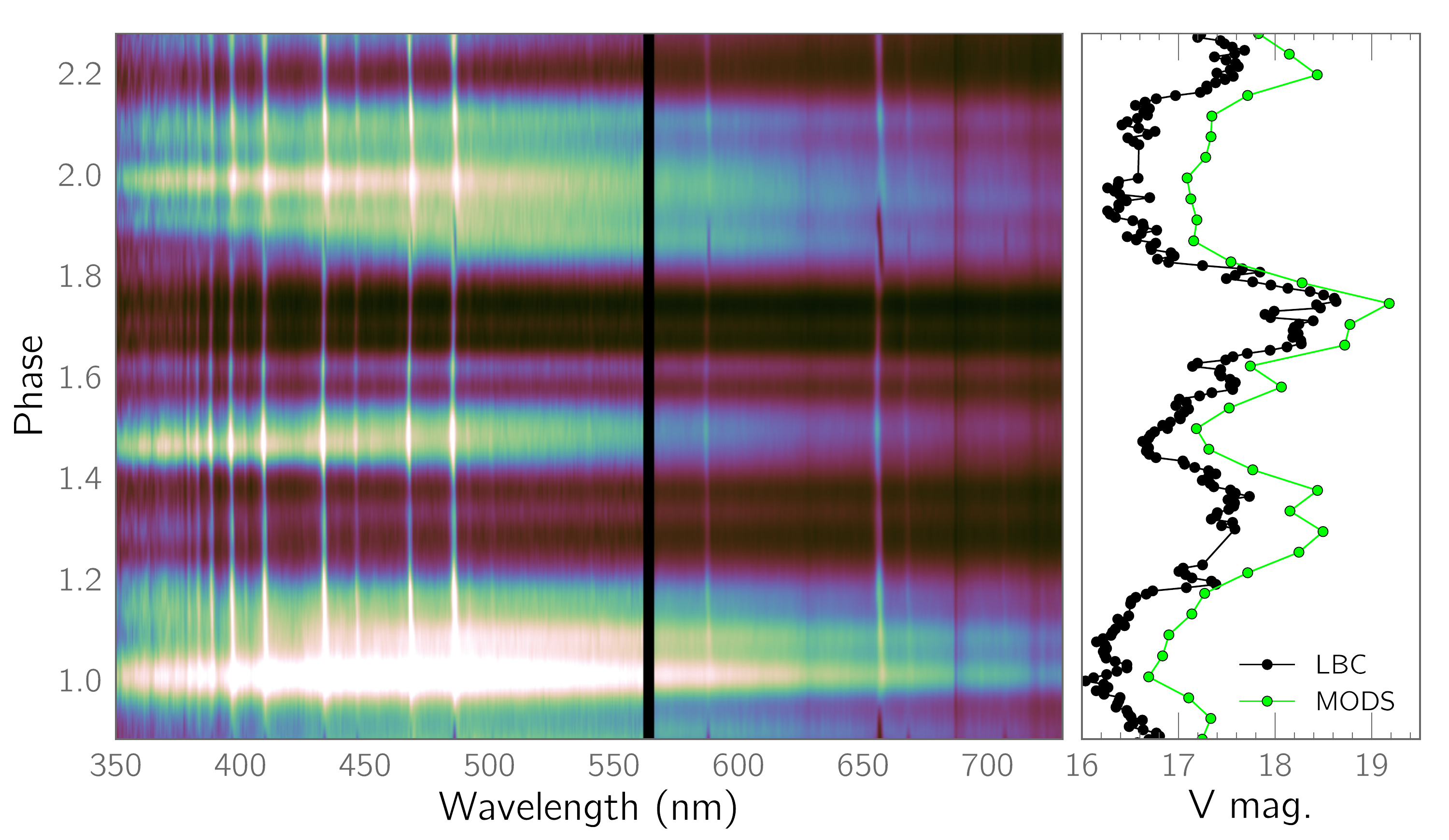}
\caption{The spectra, alongside a light curve showing the simultaneous LBC photometry and the estimated $V$ magnitudes from the flux-calibrated MODS spectra. The second half of the first absorption dip is visible at the bottom of the figure near phase 0.9, and the second dip occurs one cycle later. Note the rapidly changing continuum as J1321 rises to its both of its maxima. Telluric lines are uncorrected. \label{trailedspectrum}}
\end{figure*}

\subsection{Spectra}\label{spec}

The LBT spectra show a rich assortment of single-peaked emission lines typically seen in polars, including the Balmer series, He I, and He II $\lambda$ 4686 \AA. Both the continuum and the spectral lines undergo dramatic, orbital-phase-dependent variations, the most remarkable of which is a brief reversal of the spectral lines from emission to absorption. This phenomenon is discussed in detail in section~\ref{absorption}. 

The spectra provide insight into the variations in the optical light curve. During the photometric minimum, the continuum is weak and flat, with the Balmer jump in emission. As J1321 begins to rise from its minimum, the continuum begins to develop a smooth, prominent bulge, with the peak of the bulge moving blueward of 500~nm as the system brightens. After reaching its peak brightness, the system then fades to its secondary photometric minimum, when the bulge becomes less apparent and the continuum more closely resembles a sloped line that increases at shorter wavelengths. At the end of the secondary minimum, J1321 recovers to its secondary photometric maximum, and the continuum bulge becomes prominent again in the blue end of the spectrum. J1321 then fades back to its primary minimum, with the bulge disappearing completely. To illustrate these changes, we show our spectra and the simultaneous light curve in Fig.~\ref{trailedspectrum}.

We use cyclotron modeling in the following section to establish that the changes in the continuum are consistent with cyclotron beaming effects.

\subsection{Cyclotron Modeling}

The magnetic field of a polar can be estimated by identifying a set of cyclotron harmonic peaks or Zeeman lines in the spectra. Their absence in our data means that J1321's magnetic-field strength is difficult to directly measure. In polars with relatively low field strengths, a smooth, hump-shaped optical continuum results from the high-order cyclotron harmonics blending together. For example, in the low-field polars BL~Hyi \citep{schwope95} and LSQ1725-64 \citep{fuchs}, the field strength is $\sim$12~MG based on the detection of Zeeman lines, and each system shows a featureless cyclotron continuum in optical spectra.

In order to constrain the magnetic field strength of the WD in J1321, we use the homogeneous cyclotron emission model of \citet{chanmugam79} and \citet{wick85} to calculate optical spectra of accretion columns on a magnetic white dwarf. The shape of the spectrum depends on the plasma temperature, $kT$, and the magnetic field strength, $B$, which are assumed to be constant. The models also depend on the viewing angle, and a dimensionless depth parameter, $\Lambda$. The depth parameter is a combination of the optical path length, the magnetic field strength, and the electron density, and it determines the wavelength at which the accretion column becomes optically thin.

When bright, the spectra of J1321 show a smooth continuum with a broad peak between 400~nm and 500~nm, suggesting that we are seeing high harmonics blended together by a significant electron temperature. At plasma temperatures less than 10~keV, the homogeneous models predict that cyclotron harmonics should be noticeable in the spectra, even at fairly high harmonics. The smooth continuum of the LBT spectra implies a plasma temperature larger than 10~keV, and we adopt a temperature of 20~keV as did \citet{schwope95} for BL~Hyi.

\begin{figure}
\epsscale{1.25}
\plotone{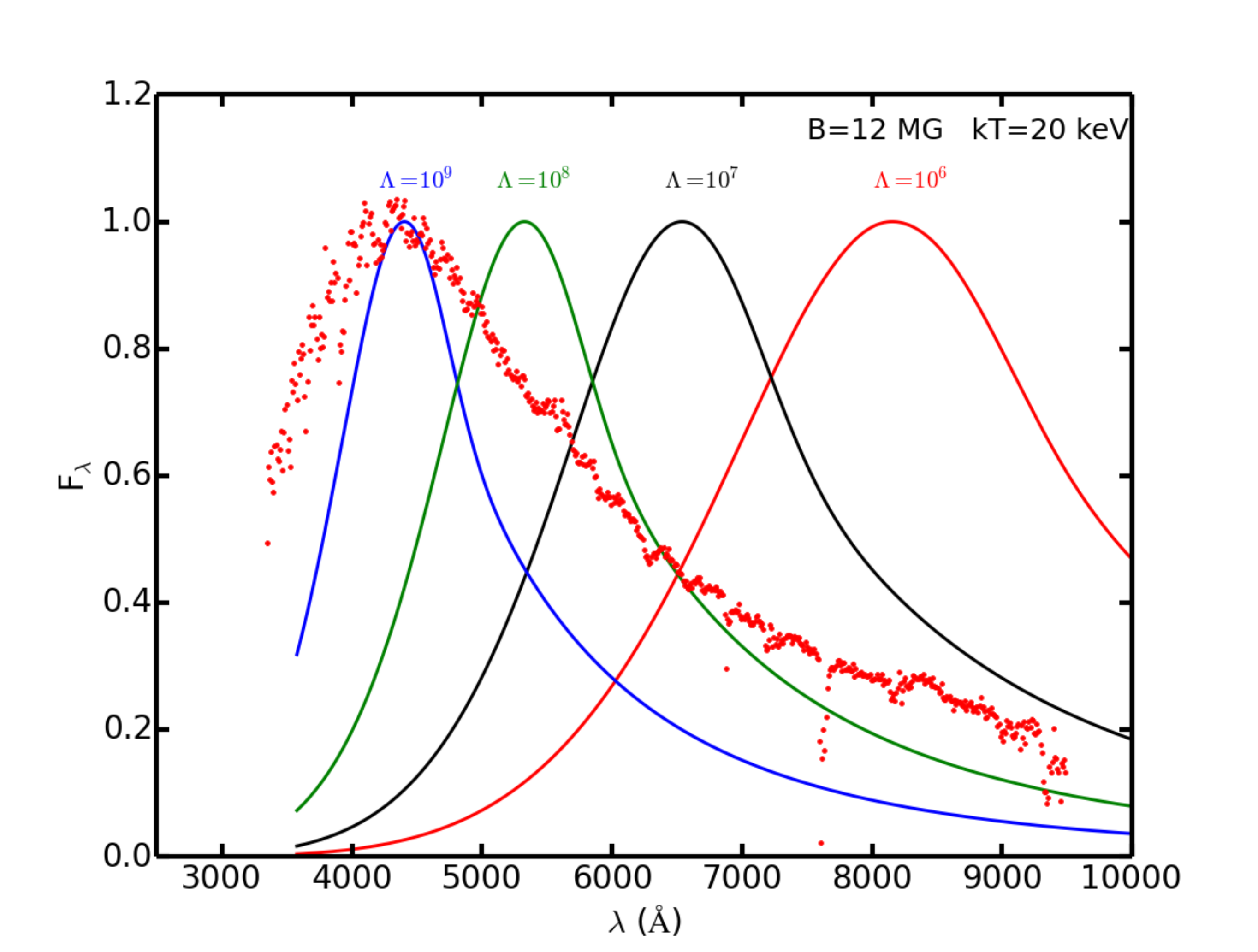}
\caption{ Homogeneous cyclotron-emission models compared with J1321's continuum (dots) from one spectrum obtained at $\phi =0.06$, close to its photometric maximum. From that spectrum, we subtracted an average spectrum of the faintest part of J1321's orbit in order to isolate the cyclotron component. Residual emission lines were removed, and the spectrum was binned into 10~\AA\ steps. \label{cyclotron}}
\end{figure}

\begin{figure}
\epsscale{1.3}
\plotone{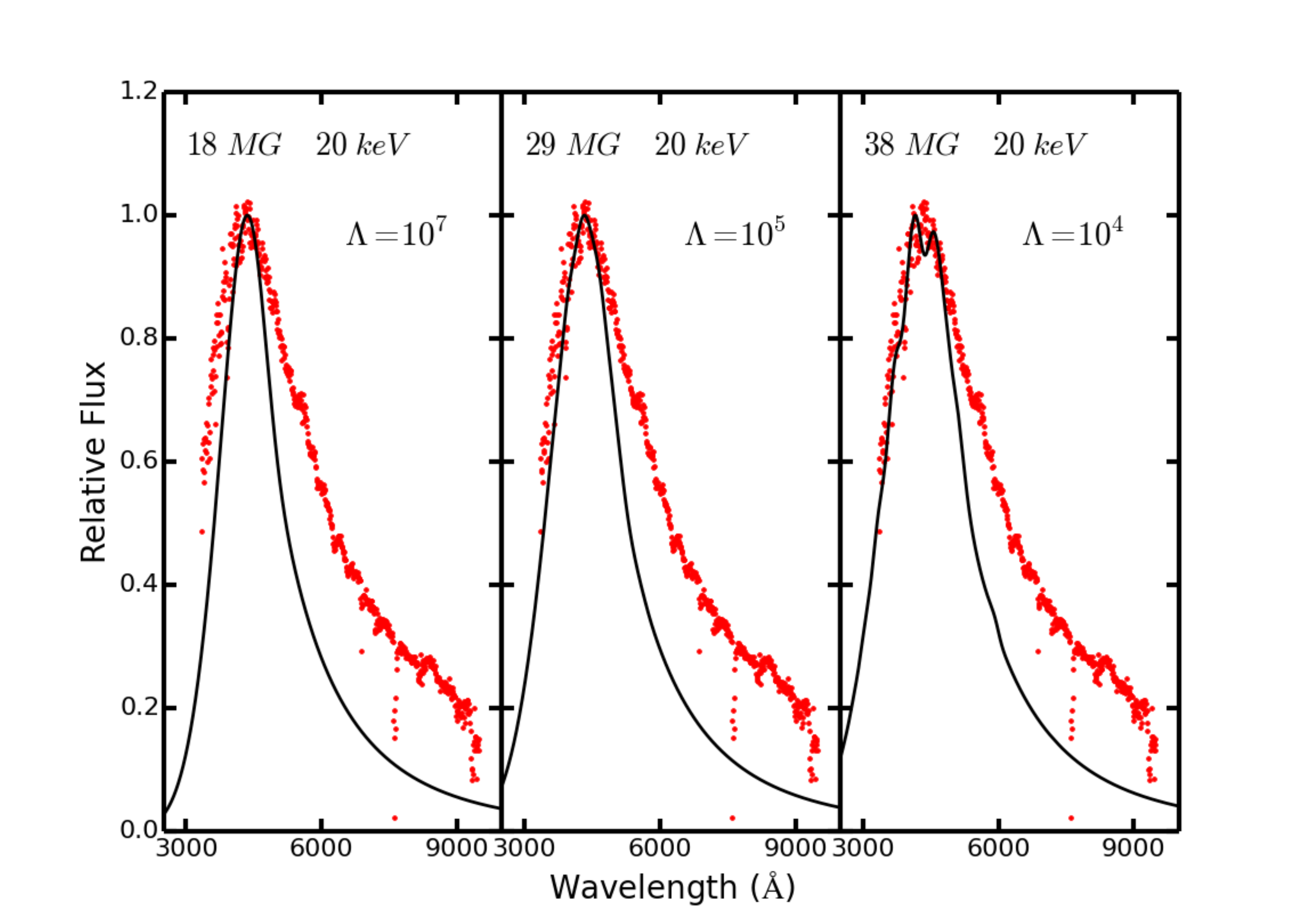}
\caption{If the plasma temperature ($kT$) is fixed at 20~keV, it is possible hold the peak wavelength of the cyclotron spectrum constant by increasing the field strength and decreasing $\Lambda$. However, at sufficiently high field strengths (here, $B$ = 38~MG), individual cyclotron harmonics begin to appear. \label{cyclotron2}}
\end{figure}

Most of the model parameters are unconstrained in J1321, so we will simply look at how the models that fit a similar system, BL~Hyi, can be adjusted to look like J1321's cyclotron spectrum at maximum light. To isolate the cyclotron spectrum for our modeling, we selected a spectrum from $\phi = 0.06$, subtracted the average spectrum of J1321 during the faintest part of its orbit, and removed the residual emission lines. The resulting cyclotron continuum peaks at 433~nm, compared to 640~nm in BL~Hyi. Adjusting the \citet{schwope95} BL~Hyi model to achieve a significant blueward shift of the peak to match J1321 requires tweaking one or more of the four free parameters of the homogeneous models. For example, as Figure~\ref{cyclotron} shows, increasing the depth parameter $\Lambda$ by two orders of magnitude to $10^9$ while keeping $kT=20$~keV and $B=12$~MG fixed will push the model peak to 433~nm. As \citet{schwope95} also noted for BL~Hyi, the model significantly underpredicts the width of the observed cyclotron spectrum.

Due to the lack of constraints on the cyclotron modeling, many different sets of parameters can produce a smooth cyclotron spectrum that peaks near 433~nm. In particular, an increase in $B$ can be substantially offset by a decrease in $\Lambda$. For example, Figure~\ref{cyclotron2} shows that if $kT = 20$~keV, the cyclotron spectrum for $B = 18$~MG and $\Lambda = 10^7$ is almost indistinguishable from the spectrum for $B = 29$~MG and $\Lambda = 10^5$, except that the latter has a slightly wider peak. Likewise, boosting the plasma temperature to $kT=35$~keV while setting $B=12$~MG and $\Lambda = 10^7$ also results in a cyclotron spectrum that peaks at the observed wavelength.

Our models fix the viewing angle at 80$^\circ$, as was done for BL~Hyi in \citet{schwope95}. Thus, there is little room to increase the viewing angle to push the peak blueward. Without knowledge of the exact orbital inclination and the magnetic colatitude of the accretion region(s), there are too many free parameters to achieve a reasonably unique fit, and the situation would be even more underconstrained if the viewing angle were allowed to vary.

Typical values for $\Lambda$ are $\sim10^5-10^7$, and it is uncommon to find $kT > 20$ keV (e.g., UZ For, \citealt{sbt90}; BL Hyi, \citealt{schwope95}; EP Dra, \citealt{sm97}). Assuming that $\Lambda$ is in this range and that $kT = 20$~keV, we estimate the field strength to be less than $\sim$30~MG. At field strengths higher than $\sim$30~MG, individual cyclotron harmonics begin to appear (Fig.~\ref{cyclotron2}) if the other model parameters are adjusted to keep the peak of the cyclotron spectrum near 433~nm.

We conclude that we can reproduce J1321's optical continuum well enough to establish that it is a cyclotron spectrum. The large residuals from the homogeneous model suggest that a more sophisticated, heterogeneous model is needed to more accurately model cyclotron spectra in polars.

\begin{figure*}
\epsscale{1.17}
\plotone{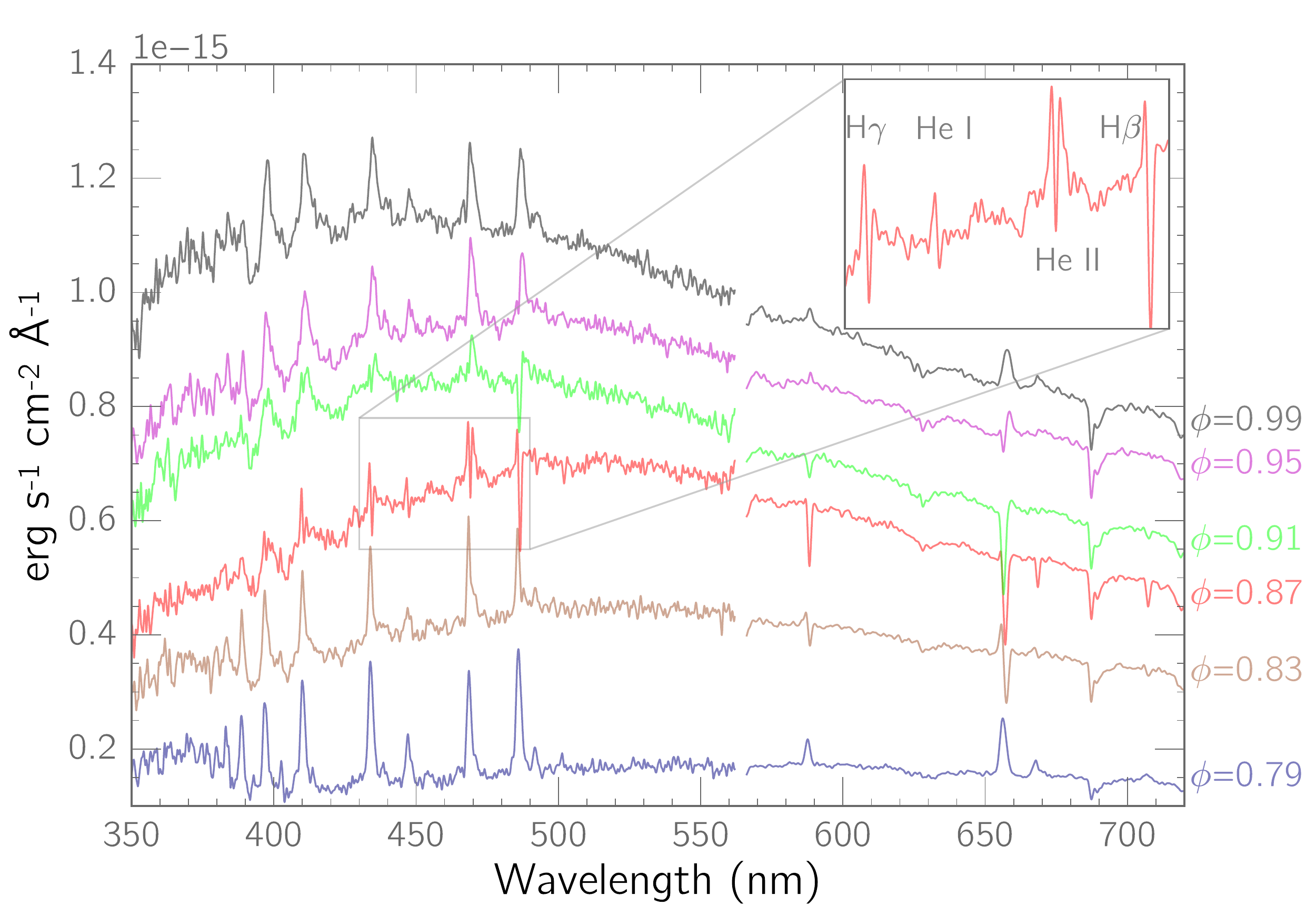}
\caption{Spectral changes in J1321 during the second absorption dip in our data, as the system rises from its minimum to its primary photometric maximum. To prevent overlap, each spectrum is offset from the previous spectrum by a constant of $0.2 \times 10^{-15}$ erg s$^{-1}$ cm$^{-2}$ \AA$^{-1}$. The absorption dip is clearly seen in H, He I, and He II. The inset plot offers a zoomed view of selected lines at the phase of maximum absorption. Additionally, the continuum develops a blueward-moving bulge as the system brightens. During the first absorption dip (not shown), more of the lines fully transitioned to absorption.\label{spectra}}
\end{figure*}

The presence of a cyclotron continuum for $\sim$85\% of the orbit implies that at least one accretion region is visible during that segment of the orbit. The absence of a cyclotron spectrum for the remainder of the orbit, without any concurrent weakening of the emission lines, points to an occultation of the accretion region(s) behind the limb of the WD. By contrast, if this were an eclipse of the cyclotron-emitting region by the secondary, there would have been at least a partial occultation of the line-forming regions. This finding accords with the fact that J1321 does not show the abrupt ingress and egress features that are almost universally observed in eclipsing polars \citep{L15}. Thus, J1321's status as a non-eclipsing system is secure, but for reasons discussed later, its orbital inclination is probably high.

\subsection{Phase-Dependent Absorption} \label{absorption}

\begin{figure*}
\epsscale{1.17}
\plotone{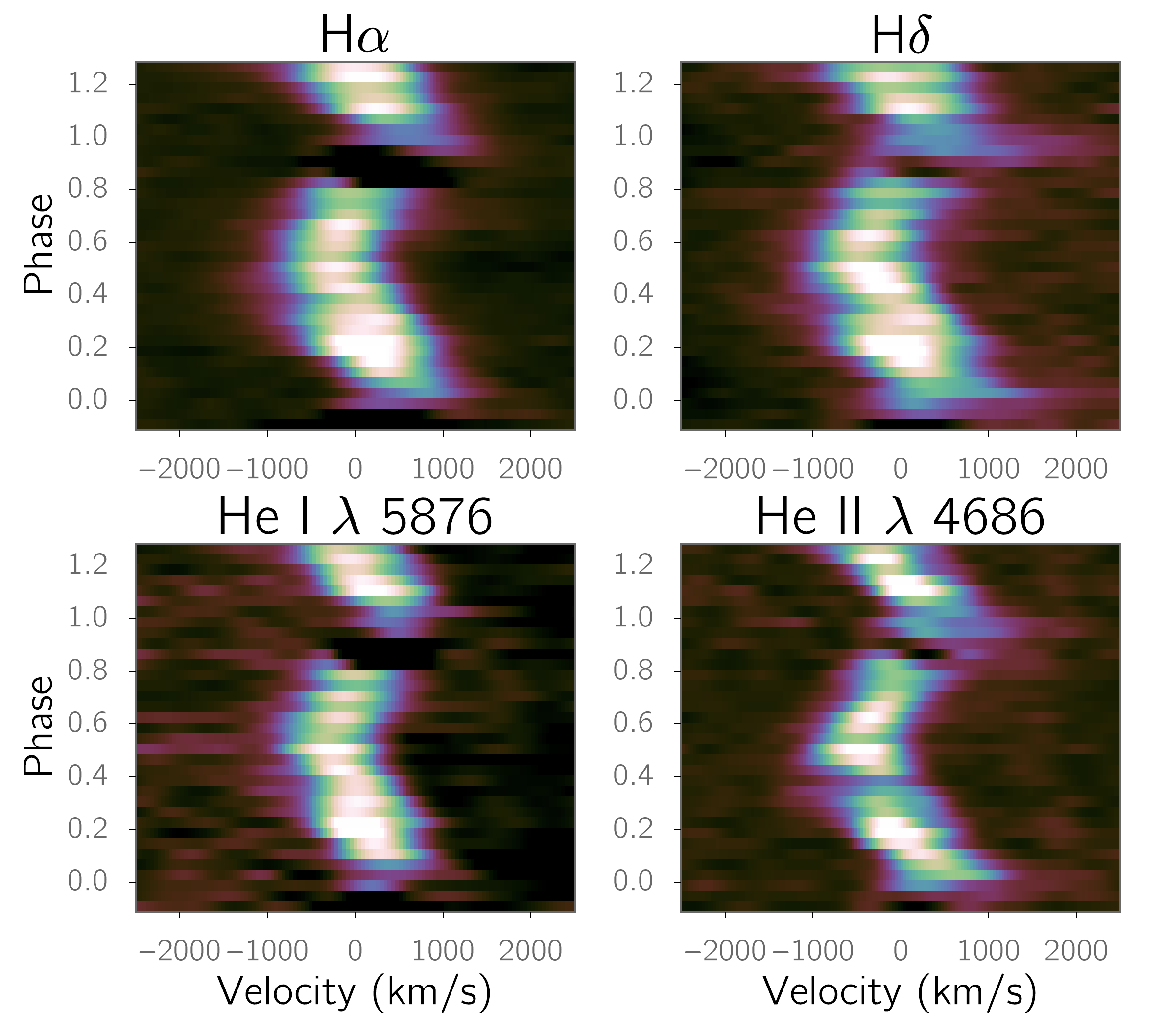}
\caption{Trailed spectra of four lines. The phases are defined such that the dip occurs at $\phi = 0.9$, as it does in other polars showing the phenomenon. As noted in the text, the systemic velocity is unknown and is assumed to be 0 km s$^{-1}$ in this figure. Near $\phi = 0.0$, high-velocity, redshifted emission is more readily seen in the higher-order Balmer lines.\label{trailedspec}}
\end{figure*}

The defining feature of J1321's spectral behavior is the appearance of deep H and He absorption lines for about one-tenth of the orbit. The LBT spectra detected two such line inversions in consecutive orbits, including the second half of the first event and the full duration of the second. As is shown in the spectra in Fig.~\ref{spectra}, for each of the optical H and He I lines, the absorption begins in the red wing of the line and quickly increases in equivalent width as it moves blueward. The emission components are quickly replaced by broad, deep absorption lines which continue to move towards shorter wavelengths. Near the end of the line inversion, the lines show P Cyg profiles with rapidly weakening absorption components. Although the wings of He II $\lambda$ 4686 \AA\ never fully transition into absorption, the line profile is bifurcated by a narrow absorption core which dips below the continuum. Like the H and He I absorption, the He II $\lambda$ 4686 \AA\ absorption component shifts blueward, but it is narrower and moves across a much smaller range of velocities. Figure~\ref{trailedspec} further illustrates this behavior with a set of trailed spectra of selected H, He I, and He II lines. As we will discuss in Section~\ref{other_polars}, emission-absorption reversals like this have been reported in only a handful of polars and are attributed to an eclipse of the accretion region by the curtain \citep{schmidt}.

We modeled the strongest absorption lines as the sum of two independent absorption Gaussians and measured the width of the absorption by numerically determining the FWHM of the combined function. The second absorption component permits the line profiles to be skewed. H$\alpha$ always had the highest FWHM, peaking at 900 km s$^{-1}$, approximately 150-200 km s$^{-1}$ higher than He I $\lambda\lambda$ 5876 \AA, 6678 \AA\ and 7066 \AA. In the blue end of the spectrum, the FWHMs for H$\beta$ and H$\gamma$ were as high as 700-800 km s$^{-1}$, compared to a peak FWHM of 600 km s$^{-1}$ for He I $\lambda$ 4471 \AA. He II $\lambda$ 4686 \AA\ was the narrowest of the absorption lines, with a FWHM of 370 km s$^{-1}$.

\begin{figure}
\epsscale{1.15}
\plotone{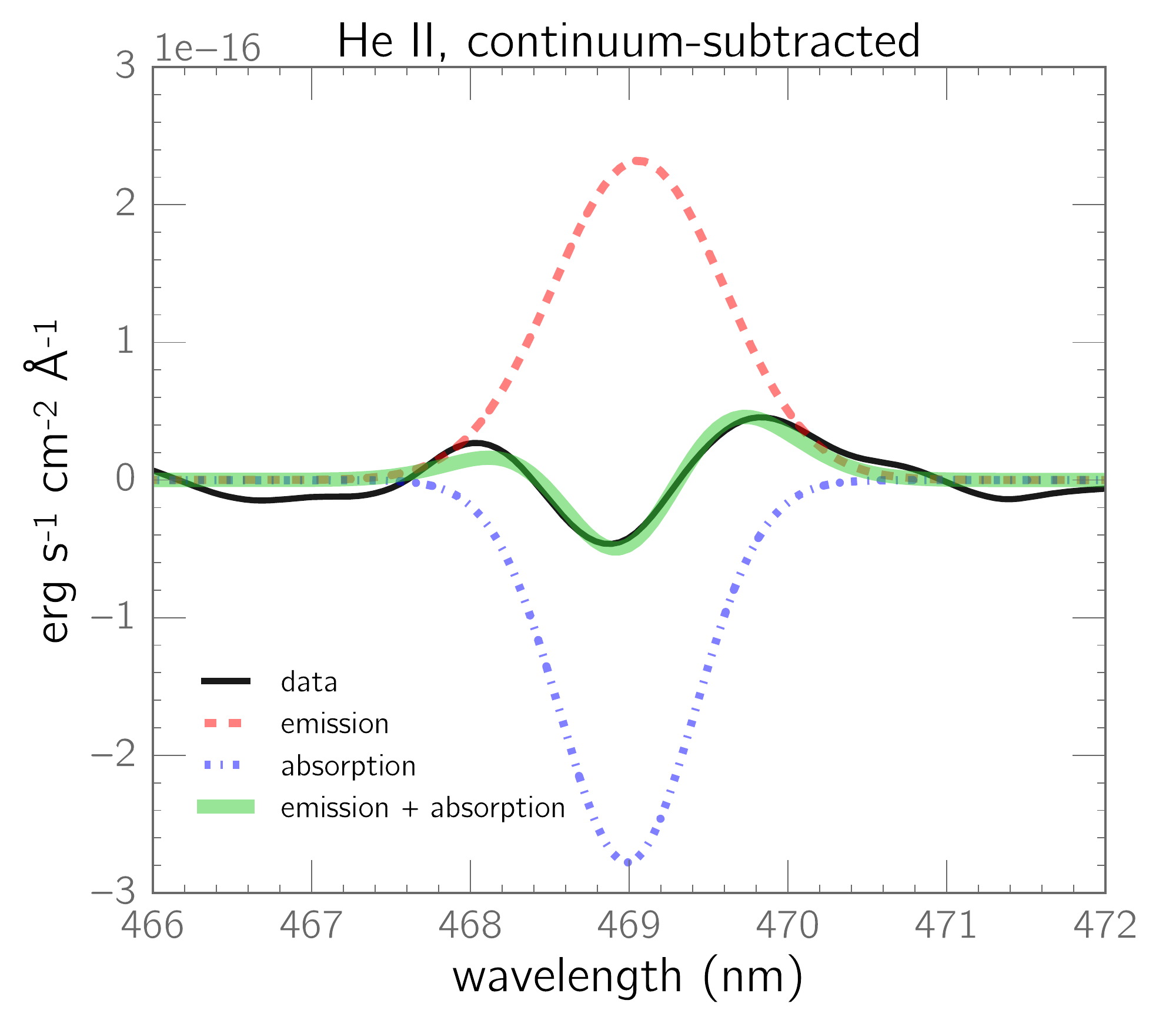}
\caption{A profile of He II $\lambda$ 468.6 nm, shown as the superposition of two Gaussians representing emission and absorption components. Contamination from the emission component causes the observed line profile to be significantly shallower and narrower than the true absorption component. The amplitude and FWHM of the emission Gaussian were fixed to their values from a spectrum taken outside of the line inversion, while all remaining parameters were allowed to vary freely. Because changes to the parameters of one of the Gaussians can be offset by changing those of the other, this fit to the data is not unique and is meant to be illustrative. \label{linemodel}}
\end{figure}

One challenge with interpreting these FWHM values is that line emission from the accretion flow probably suppresses the FWHM of the absorption, such that the observed absorption features are the sum of absorption and emission components. As an illustration, Fig.~\ref{linemodel} models He II $\lambda$ 4686~\AA\ as the sum of an emission Gaussian and an absorption Gaussian. Although the superposition of these two Gaussians yields a reasonable fit to the data, the narrow absorption feature in the observed spectrum bears little resemblance to the broad, deep absorption component in the model. The actual amplitude and FWHM of the emission component are unknown during the dip, so it is impossible to reliably recover the shape of the absorption component. If the observed absorption lines are contaminated by the presence of emission, as we suspect, then measurements of the FWHM of the observed absorption features will systematically underestimate the true FWHM of the absorption component. Thus, the FWHMs that we have reported are probably lower limits for the true absorption components.

Section~\ref{dip-discussion} explores the physical implications of the absorption event and uses it to probe the structure of the accretion curtain.


\subsection{Doppler Tomography}

\begin{figure*}
\epsscale{1.15}
\plottwo{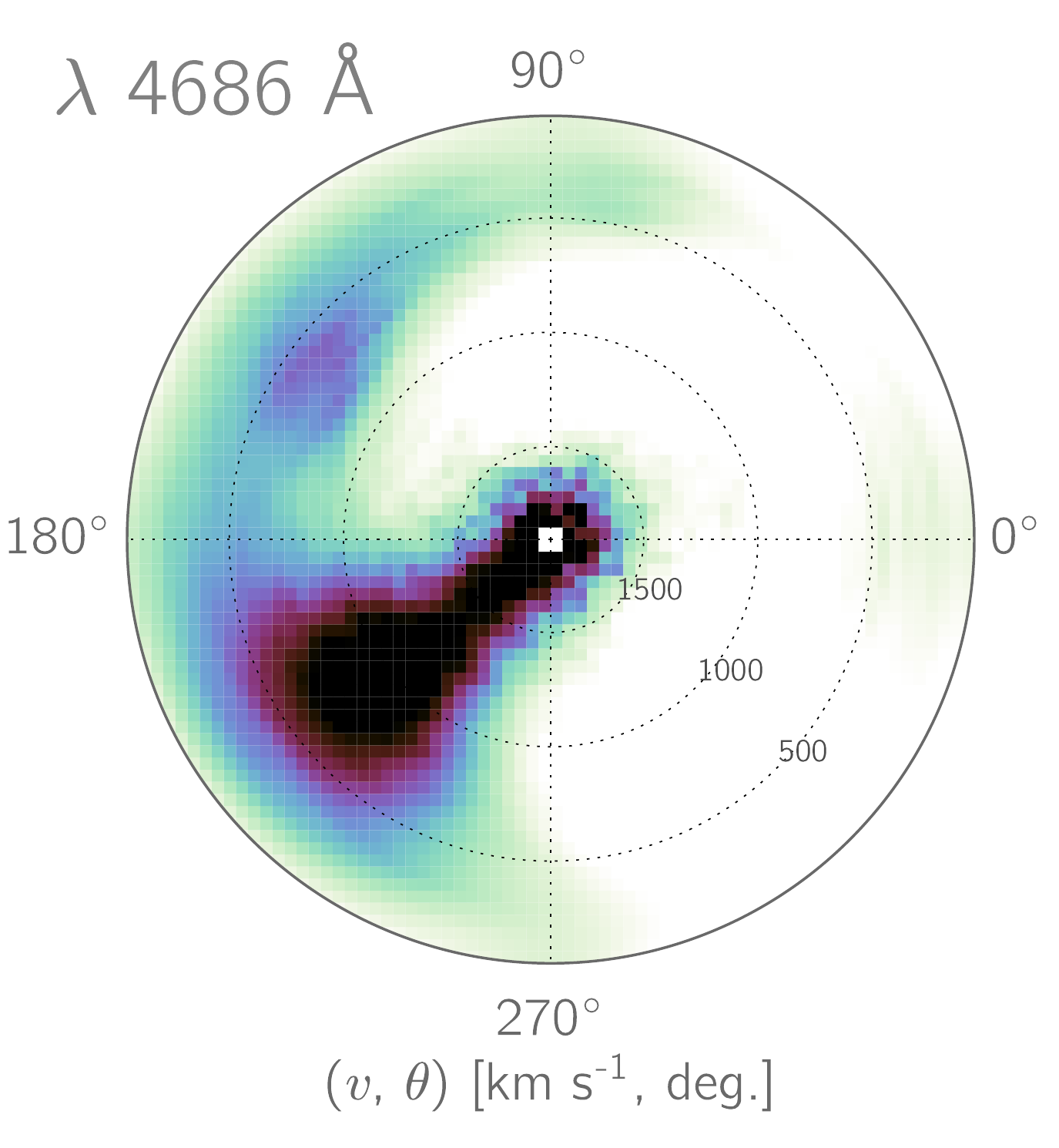}{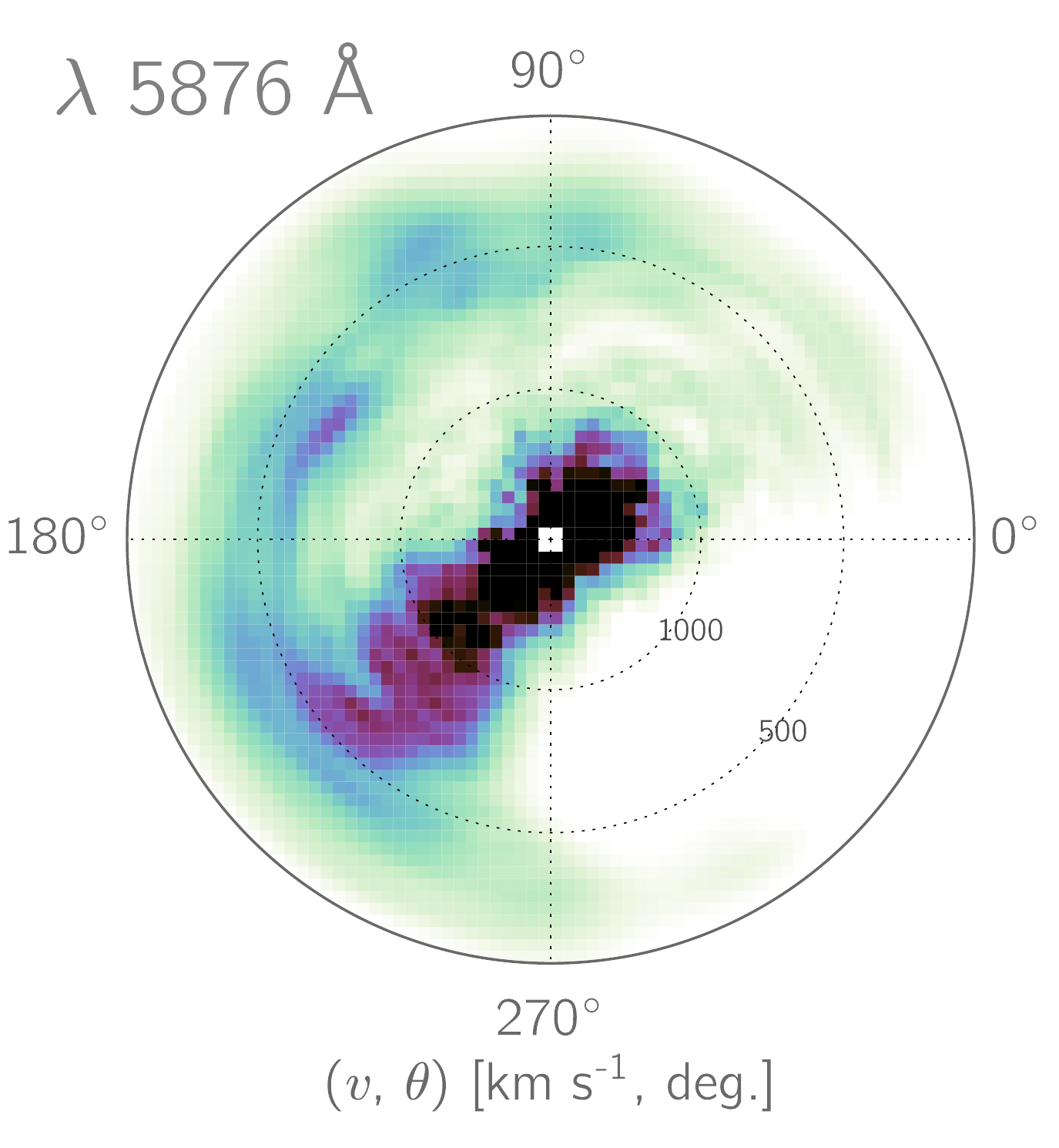}
\caption{Inside-out Doppler tomograms of He II $\lambda$~4686 \AA\ (left) and He I $\lambda$~5876 \AA\ (right). We adopt an inclination of 65$^{\circ}$ and a systemic velocity of 0 km s$^{-1}$. Because the time of inferior conjunction is unknown, the tomograms are phased such that the absorption dip is centered on $\phi\sim0.9.$ The inside-out projection clearly resolves the ballistic and magnetically-channeled regions of the accretion flow. In the He I tomogram, there is high-velocity emission near angle 45$^{\circ}$, as might be expected from radial inflow towards a second accreting pole on the opposite side of the WD. Although this emission is present in tomograms of other He I lines, it is weak in Balmer-line tomograms. \label{tomo}}
\end{figure*}

We computed Doppler tomograms of the system using the inside-out projection implemented by \citet{kotze}. By plotting the highest-velocity emission at the origin of a polar plot as opposed to its circumference, this projection excels at deblending low-velocity emission while revealing high-velocity structure. The inside-out tomograms of He~II $\lambda~4686$ \AA\ and He~I $\lambda~5876$ \AA, both of which are shown in Fig.~\ref{tomo}, clearly distinguish the ballistic portion of the stream from the accretion curtain. The absence of the donor in the tomograms and trailed spectra suggests some combination of shielding by the accretion flow and insufficient spectral resolution.

Properly phasing a tomogram requires knowledge of the time of the donor star's inferior conjunction. Since we did not detect spectral features from the donor, we estimate this parameter using two complementary methods. First, we presume that nearly all of the ballistic portion of the accretion stream produces detectable spectral lines, such that the start of the ballistic stream is a proxy for the position of the donor star. Accordingly, we adjusted the time of inferior conjunction until the stream originated near the donor's expected position above the origin. Second, absorption dips caused by the accretion stream in polars are generally observed near $\phi_{orb} \sim 0.9$, a tendency that can be used to crudely guess the time of inferior conjunction. The phasing from the two methods was similar, so it is likely that our tomograms are phased with reasonable accuracy, ensuring that they can be compared directly with those of other systems.

Another important consideration is the unknown systemic velocity, $\gamma$. As shown by \citep{schwope99}, adopting an incorrect value of $\gamma$ can introduce false structure into a Doppler map, so we tested a range of values for $\gamma$ between -200 km s$^{-1}$ and 200 km s$^{-1}$ to search for a dependence on $\gamma$. We found that the large-scale structure in our tomograms was substantially independent of $\gamma$ between $-200$ km s$^{-1}$ and 100 km s$^{-1}$, and we assumed a value of $ \gamma = 0$ km s$^{-1}$.


\section{Discussion}

\subsection{Missing Secondary Maxima}

The cause of the missing secondary maxima remains unclear, and there is no discernible pattern as to when they occur. It could be that a short-term fluctuation in the mass-transfer rate causes either a change in the extent of the cyclotron-emitting region (in the case of one-pole accretion) or a brief cessation of accretion onto a second accreting polecap (in the case of multipole accretion). While certain single-pole accretion geometries can produce a double-humped light curve because of the beamed nature of cyclotron emission, the fact that the primary maxima are unaffected by the missing secondary maxima suggests that the two maxima are produced in different regions, offering circumstantial support for two-pole accretion. DP~Leo, in which accretion onto a second pole results in a highly variable secondary photometric maximum, provides a precedent for this explanation \citep[Fig. 1 in][]{b14}.

However, any rigorous explanation of the missing secondary maxima would likely require at least some knowledge of the accretion geometry, in particular the number of accretion regions and their locations on the white dwarf. This information could be gleaned from X-ray and/or polarimetric observations. The current absence of such data prevents us from offering a more refined hypothesis for the missing maxima.

\subsection{Emission-Absorption Line Inversions}
\label{dip-discussion}

Emission-absorption line inversions occur when the accretion region is viewed through an accretion curtain that is optically thin in the continuum but optically thick in the spectral lines. Thus, our line of sight (in the emission lines) stops at the relatively cool gas in the outer curtain, thereby obscuring the underlying cyclotron continuum at the wavelengths of those transitions. In the continuum, however, the curtain has a very low optical depth, so the cyclotron emission readily passes through it \citep{verbunt}.

Emission-absorption reversals can be used to assess models of the line-producing regions of the accretion curtain. \citet{fw} modeled the H, He~I, and He~II emission of accretion funnels and found that a combination of irradiation and magnetic heating was necessary to reproduce the observed line strengths in polars. They concluded that although irradiation was sufficient to account for the H and He I fluxes, the strength of the He~II emission necessitated an additional heating mechanism. Their preferred candidate for the extra source was magnetic heating in the threading region, where the stream couples onto the magnetic field lines. Based on their modeling, the H lines are produced throughout the curtain, while the He~I and He~II emission should be mostly confined to the threading region and, to a lesser extent, the terminus of the funnel as it approaches the accretion shock.

\begin{figure}
\epsscale{1.17}
\plotone{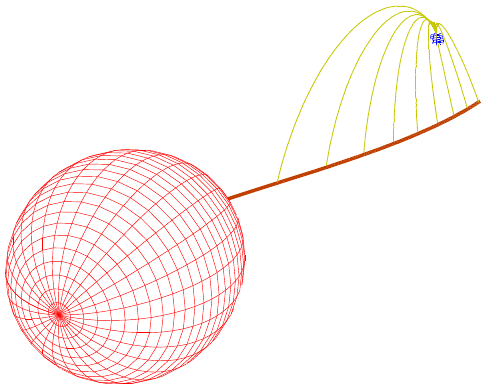}
\caption{A possible accretion geometry for J1321 shortly after the start of the line inversion, generated using the code by \citet{kotze}. The white dwarf is obscured behind the gas in the accretion curtain. Because J1321 is a non-eclipsing system, the threading region cannot pass in front of the accretion region and therefore cannot be the source of the observed He II absorption in J1321, in contrast to theory.\label{model}}
\end{figure}

Our observations support some aspects of this model while challenging others. In particular, our data are inconsistent with their prediction that He I emission is produced primarily in the threading region. It follows from the non-eclipsing nature of J1321 that the threading region---which, by definition, is confined to the orbital plane---cannot cross our line of sight to the accretion region, as is illustrated by the schematic in Fig.~\ref{model}. As a result, it cannot be the source of the He~I absorption. Furthermore, for both H and He~I, the absorption occurs across the same phases, starts in the red wing of the line, moves blueward, and fully engulfs the line during the strongest part of the dip. Thus, it is likely that the H and He~I line-forming regions are largely coextensive. The large FWHMs for both H and He~I (Sec.~\ref{absorption}) during the dip are consistent with our line of sight passing though a wide range of gas velocities within the curtain, implying that the Balmer and He~I lines are formed throughout much of the curtain. Although the $\sim$25\% higher FWHM for the Balmer lines relative to nearby He I lines shows that the corresponding line-forming regions are not entirely co-spatial, He~I must originate throughout a larger portion of the curtain than predicted by \citet{fw}.

The best interpretation of the He~II absorption is somewhat ambiguous. On one hand, the fact that H, He~I, and He~II all undergo the same blueward motion during the line inversion suggests that they share a common line-formation region within the curtain. Although the model of \citet{fw} predicts that the threading region should be the primary line-formation region for He~II, the same argument regarding the He~I absorption applies to He~II; the threading region does not cross our line of sight to J1321's accretion region and therefore cannot be the source of the He~II absorption. Consequently, the He II-absorbing material must reside within the curtain. On the other hand, the comparatively limited velocity range of the He~II absorption is compelling evidence that the He~II-emitting regions are not entirely co-spatial with the H and He~I line-forming regions.

There are at least two plausible explanations for the reduced depth and width of the He~II absorption relative to the other lines. One is that only a small region within the curtain is hot enough to produce significant He~II absorption during the line inversion, resulting in a reduced range of velocities within the absorbing gas. Alternatively, the He~II absorption might actually be deeper and wider than indicated by the narrow absorption core in our spectra, but blended emission from another He~II line-formation region could significantly dilute its appearance (see Sec.~\ref{absorption} and Fig.~\ref{linemodel}).

In spite of this ambiguity, it is evident that H and He~I are produced in similar regions throughout the curtain, while He~II is produced in at least part of the curtain. One possible explanation for these observations is that the entire curtain is significantly hotter than predicted by \citet{fw}. Unless irradiation-induced heating is stronger than predicted, this conclusion implies the existence of an additional mechanism that heats the entire curtain.

The cause of the steady blueward motion of the lines during the dip is unclear. One simple interpretation is that the threading region has a sufficiently wide azimuthal extent to intercept multiple field lines. Gas captured near one end of the threading region would therefore follow a different trajectory than gas seized at the opposite end. Thus, at the start of the dip, the absorbing gas is following field lines which move directly away from us in the radial direction; during the later phases of the dip, the gas travels along different field lines which cause it to have lower projected velocities. The significant azimuthal extent of the curtain in the tomograms ($\sim 45^{\circ}$) lends some credence to this possibility. Alternatively, as \citet{schmidt} pointed out, density variations of the curtain can cause azimuthal variations in the curtain's optical depth, impacting the radial velocity of the absorption.

There is some evidence of a secondary line eclipse exactly halfway between the two line inversions. This is most clearly seen just after phase 0.4 in the trailed spectrum for He~II from Fig.~\ref{trailedspec}, and it could be an occultation of the threading region by the curtain. If so, the non-detection of this event in the other lines would be consistent with the threading region being an enhanced source of He~II emission in comparison to the other spectral lines. However, the diminution of the He~II line is present in just one spectrum, so this possible feature should be treated skeptically unless it is confirmed by subsequent observations.

\subsubsection{Other Polars with Line Inversions}
\label{other_polars}

While emission-absorption reversals have been observed in at least four other polars, the phenomenon has unique characteristics in each of the systems. Just one other system, FL Cet \citep{schmidt}, shows the effect in all of its major lines in the optical. FL Cet is an eclipsing polar, with the dip occurring just before the eclipse. During the dip, its H and He I lines switch from emission to absorption, while He II develops a weak P Cyg absorption component. The H and He I lines behave identically throughout the absorption event and fully engulf the emission lines, exactly as is observed in J1321. \citet{schmidt} concluded, as we have for J1321, that H and He I are emitted from the same regions within the curtain. Likewise, in both systems, the He II $\lambda$ 4686~\AA\ absorption component is comparatively weak and does not fully engulf the emission.

However, the motion of the absorption in J1321 and FL~Cet is notably different. In contrast to the steady blueward motion of J1321's absorption lines, the absorption dip in FL~Cet begins in the {\it blue} wing of each line, moves redward to the line center, engulfs the entire line, and finally reverses direction, weakening as it moves blueward. Moreover, during the absorption phase, He II develops a weak P Cyg absorption component which does not move significantly, unlike the He II absorption in J1321, which appears near the line center and moves somewhat blueward.

In the other three polars with emission-absorption reversals, the effect is detectable only in a handful of lines. In EF~Eri, \citet{verbunt} observed absorption components only in the H and He I lines between 5500~\AA\ -- 7000~\AA\ (the red limit of their spectra). During the absorption phases, the profiles of the lines blueward of 5500~\AA\ are distorted but remain fully in emission. 

A similar wavelength dependence is apparent in MN~Hya \citep{rw98}. H$\alpha$ and H$\beta$ develop redshifted absorption components, and He I $\lambda$~5876~\AA\ is fully enveloped by absorption. The other H and He I lines become nearly indistinguishable from the continuum during the strongest part of the dip, but the He II $\lambda$ 4686~\AA\ line remains very prominent and is almost unaffected.

The absorption dip in the final polar, V808 Aur, is more complicated. \citet{borisov} showed that He I fully transitions into absorption, but H$\beta$ retains an emission component throughout the inversion. Its He~II $\lambda$ 4686 \AA\ line, meanwhile, shows only a partial diminution of its blue wing without ever dipping below the continuum. They reported that the absorption features moved from red to blue during the inversion.

Three of these four polars (FL~Cet, V808~Aur, MN~Hya) are eclipsing, so it appears that absorption dips are preferentially observed in high-inclination systems. If so, this favors a moderately high orbital inclination for J1321.

\subsection{Longitude of the Accretion Region}

Observations of eclipsing polars, including the three discussed in Section~\ref{other_polars}, have established that the accretion region tends to lead the donor star, such that it is viewed most directly before the orbital eclipse. However, if the orbital phases inferred from our tomograms are reasonably accurate, the accretion region is probably close to the line of centers of the donor and WD, and it might even trail the donor. As shown in Fig.~\ref{trailedspectrum}, there is no cyclotron spectrum between inferred orbital phases 0.65-0.8, implying that the accretion region is behind the limb of the WD during the phases at which it would be viewed face-on in other polars. Indeed, the appearance of the cyclotron spectrum at phase 0.8 suggests that the accretion region is located in the trailing hemisphere of the WD, but better knowledge of the orbital phases is necessary before this can be explored more rigorously.

\section{Conclusion}

Our time-resolved spectra of J1321 show that its emission lines briefly transition into absorption lines each orbit. We attribute this behavior to the occultation of the accretion region by the accretion flow. The similar behavior of the H and He I lines during the dip is consistent with these lines originating throughout the curtain, in contrast to theoretical predictions. Additionally, He II $\lambda$ 4686 also develops an absorption component, but it never fully transitions into absorption. Because J1321 is not an eclipsing system, the source of the He~II absorption cannot be the threading region and must instead be gas within the accretion curtain, implying that the curtain has a significantly higher temperature than predicted by theory.

An examination of our photometry, along with archival data, reveals that J1321 transitioned from a prolonged faint state into an ongoing and stable high state at some time between 2003 and 2007. Because the strength of the absorption undoubtedly depends in part on the mass-transfer rate, the stability of J1321's mass-transfer rate will facilitate follow-up observations of this phenomenon.

\section*{Acknowledgments}

We thank the anonymous referee for an insightful report that led to the improvement of this paper.

MRK and PG acknowledge support from the Naughton Foundation and the UCC Strategic Research Fund.

This work has been supported in part by the National Science Foundation (Grant No. PHY-1062819).

We thank Paul Mason and Joshua Santana for sharing with us their unpublished manuscript about the behavior of polars observed by CRTS.

We are grateful to R. Boyle, the Vatican Advanced Technology Telescope, and the Vatican Observatory for generous allocations of observing time.

The Digitized Sky Surveys were produced at the Space Telescope Science Institute under U.S. Government grant NAG W-2166. The images of these surveys are based on photographic data obtained using the Oschin Schmidt Telescope on Palomar Mountain and the UK Schmidt Telescope. The plates were processed into the present compressed digital form with the permission of these institutions.

The National Geographic Society - Palomar Observatory Sky Atlas (POSS-I) was made by the California Institute of Technology with grants from the National Geographic Society.

The Second Palomar Observatory Sky Survey (POSS-II) was made by the California Institute of Technology with funds from the National Science Foundation, the National Geographic Society, the Sloan Foundation, the Samuel Oschin Foundation, and the Eastman Kodak Corporation.

The Oschin Schmidt Telescope is operated by the California Institute of Technology and Palomar Observatory.

The LBT is an international collaboration among institutions in the United States, Italy and Germany. LBT Corporation partners are: The Research Corporation, on behalf of The University of Notre Dame, University of Minnesota and University of Virginia; The University of Arizona on behalf of the Arizona university system; Istituto Nazionale di Astrofisica, Italy; LBT Beteiligungsgesellschaft, Germany, representing the Max-Planck Society, the Astrophysical Institute Potsdam, and Heidelberg University; and The Ohio State University.

This paper used data obtained with the MODS spectrographs built with funding from NSF grant AST-9987045 and the NSF Telescope System Instrumentation Program (TSIP), with additional funds from the Ohio Board of Regents and the Ohio State University Office of Research.


\end{document}